# Bright-field microscopy of transparent objects: a ray tracing approach

A. K. Khitrin[1] and M. A. Model[2*]

[1]Department of Chemistry and Biochemistry, Kent State University, Kent, OH 44242
[2]Department of Biological Sciences, Kent State University, Kent, OH 44242
[*]Corresponding author: mmodel@kent.edu

**Abstract**
Formation of a bright-field microscopic image of a transparent phase object is described in terms of elementary geometrical optics. Our approach is based on the premise that image replicates the intensity distribution (real or virtual) at the front focal plane of the objective. The task is therefore reduced to finding the change in intensity at the focal plane caused by the object. This can be done by ray tracing complemented with the requirement of conservation of the number of rays. Despite major simplifications involved in such an analysis, it reproduces some results from the paraxial wave theory. Additionally, our analysis suggests two ways of extracting quantitative phase information from bright-field images: by vertically shifting the focal plane (the approach used in the transport-of-intensity analysis) or by varying the angle of illumination. In principle, information thus obtained should allow reconstruction of the object morphology.

**Introduction**
The diffraction theory of image formation developed by Ernst Abbe in the 19th century remains central to understanding transmission microscopy (Born and Wolf, 1970). It has been less appreciated that certain effects in transmission imaging can be adequately described by geometrical, or ray optics. In particular, the geometrical approach is valid when one is interested in features significantly larger than the wavelength. Examples of geometrical description include explanation of Becke lines at the boundary of two media with different refractive indices (Faust, 1955) or the axial scaling effect (Visser et al, 1992). In this work, we show that a model based on geometrical optics can be used for describing the formation of a brightfield transmission image of a refractive (phase) specimen.

Our approach is based on the notion that in an infinite tube length microscope, an image replicates the real or virtual intensity distribution at the front focal plane of the objective. Thus, the effect of a refractive object can be analyzed by examining the pattern formed by extending the incoming rays back to the focal plane. Figure 1 illustrates the concept. In the absence of a specimen, the intensity distribution at the focal plane is uniform, and no image is formed. A specimen causes turning of the rays, which is equivalent to having lighter and darker areas at the focal plane. These darker and lighter areas are translated into the image.

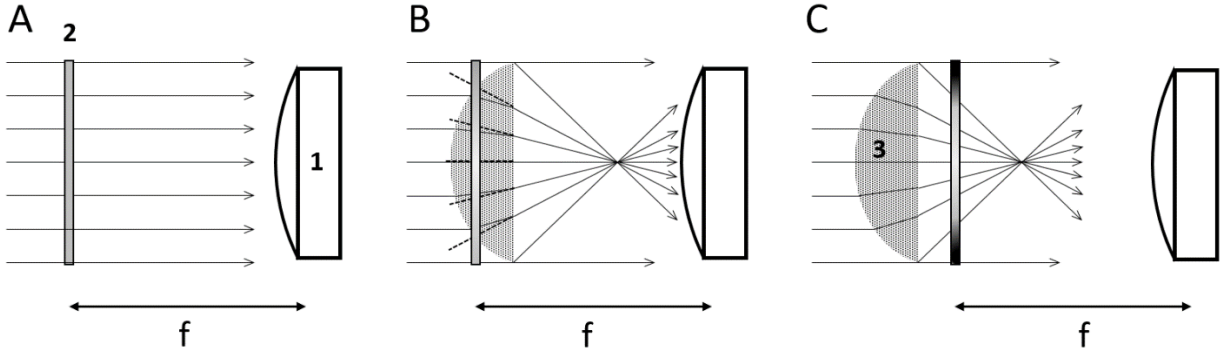

*Fig. 1.* In the absence of a specimen, a uniform illumination at the focal plane (2) of the objective (1) produces no contrast (A). The refractive specimen (3), which, for the sake of simplicity, is depicted here as a lens, alters the distribution of intensity at the focal plane; the latter may lie inside (B) or outside (C) the sample. The intensity pattern at the focal plane is reproduced by the optical system and generates an image. In the case (B), one expects a lighter area in the central part, where the density of back-projected rays is higher. In the case (C), the area immediately outside the cone of light formed by the specimen is completely dark. (This effect can be easily observed by putting a lens under the microscope).

**Theory and Discussion**
Next, we present the above model of image formation in quantitative terms. Consider a typical situation in light microscopy (Figure 2), where an object (e.g., a biological cell) is attached to the coverglass on the side of the objective. It is illuminated by light coming from the condenser on the opposite side. The cell has a slightly higher refractive index than the surrounding liquid (typically by ~2-3%) and is assumed to have a homogeneous structure. The focal plane of the objective is positioned approximately on the level of the cells but can be shifted up or down by moving either the sample or the objective.

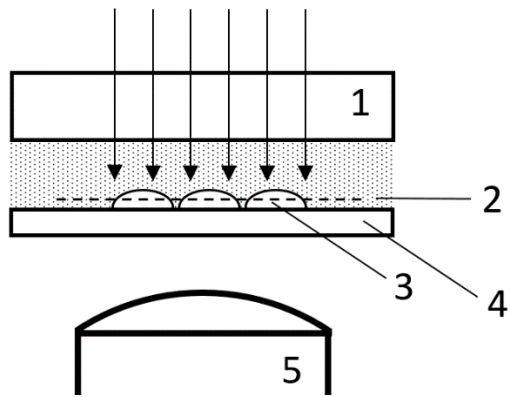

*Fig. 2. General setup in transmission brightfield imaging. 1 - slide; 2 – focal plane; 3 – object; 4 – coverglass; 5 – objective.*

Figure 3 gives a more detailed view of the ray path through the sample. A single refraction at the interface of the cell and its aqueous environment is assumed. It causes a change in the distribution of intensity at the focal plane of the objective, which, in turn, determines the intensity distribution at the image plane. It is possible, of course, to have a situation when the focal plane is below the cell, in which case the intensity would also be affected by a second refraction at the cell-coverglass boundary. This case will not be considered.

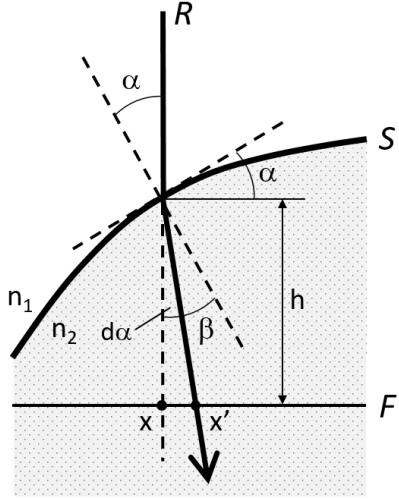

*Fig. 3.* The diagram of rays crossing the sample. A vertical incident ray $R$ impinges on the object boundary S at the angle α. Refraction causes its deflection from the vertical line by dα. The point of intersection of $R$ and $S$ lies at the distance h from the focal plane $F$. As a result, the point where the $R$ ray crosses the focal plane $F$ shifts from x to x'. The focal plane is designated as xy, and the refraction takes place in the xz-plane.

Figure 3 introduces the main parameters used in the model. From the law of refraction, we have

$$\frac{\sin \alpha}{\sin \beta} = \frac{n_2}{n_1} = n, \qquad (1)$$

where n is the relative refractive index. To simplify the following derivations, we assume that n - 1 << 1, which is usually true for live biological cells. Then, the angle of refraction dα = β - α is small, and Eq. (1) can be written as

$$\frac{\sin \alpha}{\sin(\alpha - d\alpha)} = \frac{\sin \alpha}{\sin \alpha - d\sin \alpha} = \frac{\sin \alpha}{\sin \alpha - d\alpha \cos \alpha} = \frac{1}{1 - d\alpha/\tan \alpha} = n. \qquad (2)$$

From here we find the refraction angle:

$$d\alpha = (n-1)\tan \alpha = (n-1)\frac{dh(x)}{dx}, \qquad (3)$$

where h(x) is the height of the object boundary over the focal plane (it can be positive or negative), and x is the position on the focal plane where it would be intersected by the incident ray in the absence of the sample. Refraction causes a shift of the intersection point from x to x'. For small refraction angles dα,

$$x' = x + h(x)d\alpha = x + h(x)(n-1)\frac{dh(x)}{dx}. \qquad (4)$$

Image intensity I(x') satisfies the equation
$$I(x')dx' = I_0(x)dx, \qquad (5)$$

where $I_0(x)$ is the incident light intensity. Eq. (5) states the conservation of the number of rays: all the incident rays that would arrive at the element dx without refraction are deflected to the element dx' in the presence of refraction. If we assume uniform illumination and use relative intensities, we can set $I_0(x) = 1$. Then,

$$I(x') = \frac{dx}{dx'} = \frac{1}{dx'/dx}, \qquad (6)$$

or, taking Eq. (4) into account,

$$I(x') = [1 + (n-1)\{h(x)\frac{d^2h(x)}{dx^2} + \left(\frac{dh(x)}{dx}\right)^2\}]^{-1}. \tag{7}$$

2D generalization can be done as follows. By analogy with Eq. (4), the position vector (x',y') on the focal plane, where the continuation of refracted rays crosses the focal plane, can be written as

$$(x', y') = (x, y) + h(x, y)(n-1)\, \nabla h(x, y), \tag{8}$$

where $\nabla$ is the 2D gradient in the xy-plane. In terms of the components,

$$x' = x + h(x,y)\,(n-1)\frac{\partial h(x,y)}{\partial x}, \quad y' = y + h(x,y)\,(n-1)\frac{\partial h(x,y)}{\partial y}. \tag{9}$$

The line elements dx and dx′ in Eq. (6) have to be replaced by the area elements dxdy and dx′dy′, and the equivalent of Eq. (6) takes the form

$$I^{-1}(x', y') = \left|\frac{\partial(x',y')}{\partial(x,y)}\right|. \tag{10}$$

Thus, the inverse intensity of the image is the Jacobian determinant $\left|\frac{\partial(x',y')}{\partial(x,y)}\right|$. By substituting x′ and y′ from Eq. (9), one obtains

$$I^{-1}(x', y') = \frac{\partial x'}{\partial x}\frac{\partial y'}{\partial y} - \frac{\partial y'}{\partial x}\frac{\partial x'}{\partial y} =$$
$$= [1 + (n-1)\{h\frac{\partial^2 h}{\partial x^2} + \left(\frac{\partial h}{\partial x}\right)^2\}][1 + (n-1)\{h\frac{\partial^2 h}{\partial y^2} + \left(\frac{\partial h}{\partial y}\right)^2\}] \tag{11}$$
$$- [(n-1)\left(h\frac{\partial^2 h}{\partial x \partial y} + \frac{\partial h}{\partial x}\frac{\partial h}{\partial y}\right)]^2.$$

This equation can be simplified if we assume that variations of the image intensity are small compared to the average intensity. Then,

$$I^{-1}(x', y') = 1 + (n-1)\{h\left(\frac{\partial^2 h}{\partial x^2} + \frac{\partial^2 h}{\partial y^2}\right) + \left(\frac{\partial h}{\partial x}\right)^2 + \left(\frac{\partial h}{\partial y}\right)^2\}. \tag{12}$$

This can be viewed as a general equation for image intensity. Because its right part contains the first and the second derivatives of the object profile, the interpretation of image intensity in terms of Eq. 12 is not straightforward. It is possible, however, to create a meaningful contrast from several images taken under slightly different conditions.

*1) Shift of the focal plane*

A shift of the focal plane by dz corresponds to the substitution h(x,y) → h(x,y) − dz in Eq. (12) and has no effect on the derivatives. Therefore, one can construct the difference

$$I_2^{-1}(x', y') - I_1^{-1}(x', y') = -dz\,(n-1)\left(\frac{\partial^2}{\partial x^2} + \frac{\partial^2}{\partial x^2}\right)h(x, y), \tag{13}$$

where $I_1$ and $I_2$ are the image intensities at two different positions of the focal plane. Now Eq. (13) has a simple interpretation: the contrast is proportional to the local curvature of the object boundary. This is a well-known fact used, for example, in "defocusing" microscopy (Agero et al, 2004).

It is interesting to compare this result with the so-called transport of intensity equation (TIE) (Teague, 1983; Streibl, 1984) obtained from the paraxial wave equation. The TIE equation for the "logarithmic derivative" (Equation 7b in Streibl (1984) at uniform transmittance) is

$$\left(\frac{d}{dz}\right) \ln I(x, y; z = 0) = -\Delta \varphi(x, y), \tag{14}$$

where $\Delta$ is the 2D Laplacian and $\varphi(x,y)$ is the phase. If we use a low-contrast approximation (as in Eq. (12)) and realize that (n-1)h(x,y) is equivalent to the phase $\phi(x,y)$, then Eq. (13) and Eq. (14) become identical. Both are the Poisson equations for the object profile h(x,y), where the left side represents experimental data. In the presence of image noise and for objects with complex shape, this equation is difficult to solve. Indeed, by applying the 2D Gauss theorem, one can see that perturbation from a noisy pixel does not decay, but grows logarithmically with the distance from the pixel. Within the TIE approach, various computational methods have been developed to minimize artifacts in restored phase maps [Paganin and Nugent, 1998; Volkov et al, 2002; Petruccelli et al, 2013; Waller et al, 2010).

In practical realization of the vertical shift method, two additional considerations apply. First, the vertical shift of the focal plane is not equivalent to the vertical shift of the objective (or of the stage), which is the only distance reported by the microscope hardware. Thus, to obtain the true dz to be used in the calculations, the nominal shift must be multiplied by the ratio of $n_1$ (or $n_2$, since they are nearly equal) to the refractive index of the immersion medium of the objective (Visser et al, 1992; Carlsson, 1991). Second, the focal plane must remain within the sample or the medium, but not within the coverglass, as that would introduce a second refraction which is not accounted for by Eq. (13). For example, if the first image is focused on the coverglass surface, the second or subsequent images should be focused further into the sample.

*2) Varying the illumination angle.*

The other way of creating an interpretable contrast is to vary the illumination angle, for example, by using an off-center condenser diaphragm. Variable angle illumination has been used in differential phase contrast microscopy (Hamilton and Sheppard, 1984; Chen et al, 2016) and computer tomography (Sung et al, 2009). Here we show that quantitative data can in principle be extracted from the above ray model. If $\gamma_x$ and $\gamma_y$ are small tilt angles in the xz- and yz-planes, one can use two pairs of images, taken at $\pm \gamma$. Eq. (3) is modified as

$$d\alpha = (n-1)\tan(\alpha + \gamma) = (n-1)\frac{\tan\alpha + \tan\gamma}{1 - \tan\alpha \tan\gamma} \approx (n-1)\frac{\tan\alpha + \gamma}{1 - \gamma\tan\alpha}. \tag{15}$$

At $\gamma \tan \alpha \ll 1$, the difference between each pair of images can be expressed as

$$I_{+\gamma x}^{-1}(x',y') - I_{-\gamma x}^{-1}(x',y') = 4\gamma_x(n-1)\frac{\partial h}{\partial x}\left[1 + \left(\frac{\partial h}{\partial x}\right)^2\right],$$

$$I_{+\gamma y}^{-1}(x',y') - I_{-\gamma y}^{-1}(x',y') = 4\gamma_y(n-1)\frac{\partial h}{\partial y}\left[1 + \left(\frac{\partial h}{\partial y}\right)^2\right]. \quad (16)$$

These are algebraic equations for $\frac{\partial h}{\partial x}$ and $\frac{\partial h}{\partial y}$. When the derivatives are not too large, the contrast essentially represents the slope of the profile h(x,y) along the corresponding direction. After the derivatives are found, the profile h(x,y) can be obtained by simple integration. However, one complication might arise if one strives for a higher resolution. The images forming each pair in Eq. (16) are misregistered by the amount γ h(x,y). This makes the task of numerical reconstruction of the profile h(x,y) at high resolution less straightforward.

In summary, we have presented a simple theory of transmission image formation based on ray tracing. The theory relates directly to the quantity of interest – the object profile. If the profile of a cell is known from independent measurements, one should be able to find the average refractive index, as well as related quantities – water and protein concentration. Local protein/water variations are usually less important than the integral values over the entire cell volume, and thus the geometrical description is appropriate. Although ray tracing is a very simplified description of light propagation, our results are equivalent to those based on paraxial wave theory (Teague, 1983; Streibl, 1984). The other finding is the possibility of extracting quantitative phase information from variable illumination angle, which leads to simpler equations. Future work will test the practicality of this approach.

**Acknowledgements**

The research was supported by the University Research Council Grant to MM. AK was supported by the ACS Petroleum Research Fund 58813-ND6.